\documentclass[pre,aps,preprint,amsmath,superscriptaddress]{revtex4}
\usepackage[utf8]{inputenc}
\usepackage{indentfirst}
\usepackage{dcolumn}
\usepackage{bm}
\usepackage{cases}
\usepackage{graphicx}
\usepackage{amsmath}
\usepackage{siunitx}
\usepackage{amssymb}
\usepackage{textcomp}
\usepackage{gensymb}

\begin{document}

\title{Unidirectional zero-index and omnidirectional hybrid hydrodynamic cloaks constructed from isotropic media with anisotropic geometry}

\author{Gaole Dai} \thanks{gldai@ntu.edu.cn}
\affiliation{School of Physical Science and Technology, Nantong University, Nantong 226019, China}

\author{Yuhong Zhou}
\affiliation{Department of Physics, State Key Laboratory of Surface Physics, and Key Laboratory of Micro and Nano Photonic Structures (MOE), Fudan University, Shanghai 200438, China}

\author{Jun Wang} 
\affiliation{School of Energy and Materials, Shanghai Polytechnic University, Shanghai 201209, China}
	\affiliation{Shanghai Engineering Research Center of Advanced Thermal Functional Materials, Shanghai 201209, China}

\author{Zhuo Li} 
\affiliation{School of Physical Science and Technology, Nantong University, Nantong 226019, China}

\author{Jinrong Liu} 
\affiliation{Department of Physics, State Key Laboratory of Surface Physics, and Key Laboratory of Micro and Nano Photonic Structures (MOE), Fudan University, Shanghai 200438, China}

	\author{Fubao Yang}\thanks{byang@gscaep.ac.cn}
\affiliation{Graduate School of China Academy of Engineering Physics, Beijing 100193, China}
\affiliation{Department of Physics, State Key Laboratory of Surface Physics, and Key Laboratory of Micro and Nano Photonic Structures (MOE), Fudan University, Shanghai 200438, China}

\author{Jiping Huang}\thanks{jphuang@fudan.edu.cn}
\affiliation{Department of Physics, State Key Laboratory of Surface Physics, and Key Laboratory of Micro and Nano Photonic Structures (MOE), Fudan University, Shanghai 200438, China}

\date{May 20, 2025}

\begin{abstract}
Hydrodynamic cloaking offers a promising approach for manipulating viscous flows by redirecting fluid around an obstacle without inducing external disturbances. By extending pseudo-conformal mappings into potential flow models, we introduce a new isobaric boundary condition that enables the construction of zero-index cloaks using isotropic and homogeneous media shaped into anisotropic geometries, such as elliptical shells. Compared to conventional cloaks, which suffer performance degradation under realistic viscous conditions, the zero-index design significantly reduces such losses by suppressing flow disturbances at the inner boundary. To overcome practical limitations in realizing ideal isobaric conditions, we further propose a hybrid cloak that integrates a raised fluid domain with an auxiliary flow channel above the obstacle. This architecture removes the need for viscosity tuning and, under anisotropic geometries, surpasses both conventional and zero-index cloaks in omnidirectional performance. The design is validated through simulations and experiments. Our findings offer a generalizable strategy for controlling viscous flows and open new directions for microfluidic applications including drug delivery, particle steering, and cell sorting.

\end{abstract}

\maketitle

\section{Introduction}

Fluid dynamics is a fundamental physical phenomenon common in both nature and industry, and controlling fluid motion is of broad scientific and technological interest.  The emerging field of hydrodynamic metamaterials~\cite{Morton2008,Chen2023,Xu2024,Wang2025} explores the use of engineered microstructures to achieve functionalities beyond those of conventional methods.
Among the various strategies, transformation optics~\cite{Pendry2006,Leonhardt2006}, originally developed in controlling electromagnetic waves, has become the dominant framework~\cite{Urzhumov2011,Park2019,Chen2022,Pang2022,Dai2023,Pang2024}. A key application is fluid cloaking or shielding~\cite{Urzhumov2011,Park2019,Chen2022,Pang2022,Dai2023}, which aims to hide an obstacle from the surrounding flow field. Designing an invisible cloak requires a topologically non-homeomorphic transformation that creates a hole (the cloaked region) in the physical space, thereby preventing field penetration. Meanwhile, the outer boundary remains unchanged to ensure impedance matching with the surrounding medium. This eliminates external disturbances and renders the cloak effectively invisible.

However, several fundamental challenges hinder further progress in this area. Unlike electromagnetic waves, the Navier–Stokes equations are not form-invariant under cloak transformations due to the presence of viscous and inertial terms~\cite{Dai2023}. As a result, transformation optics can only be applied to idealized or approximate cases.
Surface waves in inviscid fluids were studied earlier as an independent topic because they satisfy a Helmholtz-type equation, sharing the same form as the Maxwell equations for transverse magnetic modes~\cite{Zhu2024}.
In contrast, controlling viscous flow, as we study here, requires adherence to potential-flow-like models such as Darcy’s law~\cite{Urzhumov2011} or the Hele-Shaw cell~\cite{Dai2023,Boyko2021,Boyadjian2022,Tay2022,Chen2024}. 
Even in these regimes, most prior work focused only on bulk material properties, often overlooking the critical role of boundary conditions on the newly formed inner surface of the cloak. Since cloaking transformations correspond to mappings between boundary-value problems~\cite{Yavari2019}, ignoring such boundaries undermines both the rigor and completeness of the design.
Furthermore, designs based on non-conformal transformations, such as those producing annular isotropic cloaks, often yield strongly anisotropic and inhomogeneous parameters, especially in viscosity~\cite{Park2019}. These stringent material requirements pose practical difficulties. While conformal mappings preserve isotropy, they generally fail to match boundary impedance, which is essential for achieving true invisibility. Therefore, developing a general design framework that avoids medium anisotropy remains an urgent and fundamental challenge.

In this work, we extend pseudo-conformal mappings, originally developed for Fickian diffusion~\cite{Dai2023b}, into potential flow models to achieve hydrodynamic cloaking using isotropic and homogeneous media. We show that under the same transformation, the direction of the applied pressure bias dictates the boundary condition on the inner surface of the cloak, giving rise to two types of directional cloaks: one with a no-penetration boundary and the other with an isobaric boundary. The latter corresponds to an unconventional design in which the inner surface mimics the behavior of zero-index media in optics or thermotics~\cite{Liberal2017,Kinsey2019,Li2019,Xu2020}.
In more realistic scenarios, such as viscous flows in a finite Hele-Shaw cell, the no-penetration condition must be replaced by the no-slip condition. This substitution significantly degrades the performance of conventional cloaks, whereas zero-index cloaks remain more robust. Nevertheless, imposing an isobaric condition remains technically challenging.
To overcome this, we combine the design principles of both cloaks by increasing the depth of the cloak  and incorporating an additional flow channel above the obstacle. This hybrid design circumvents the need for viscosity-based regulation by exploiting a grooved surface geometry. Remarkably, the resulting hybrid cloak achieves superior omnidirectional performance compared to both conventional and zero-index cloaks. These results are validated by numerical simulations and experimental measurements.

\section{Zero-index  metamaterial-free hydrodynamic cloak}

To apply transformation optics, we consider a potential flow model governed by the following equations
\begin{equation} 
	\left\{\begin{array}{l}\nabla\cdot\mathbf v=0,\\ 	\mathbf v=-\alpha \nabla p,\end{array}\right.
\end{equation}
where $\mathbf{v}$ is the velocity and $p$ is the static pressure.
This model describes either creeping flow in a porous medium governed by Darcy's law~\cite{Urzhumov2011} or Hele-Shaw~\cite{Dai2023} flow between closely spaced plates, where the parameter $\alpha$ represents either $\frac{k}{\mu}$ (with $k$ the permeability and $\mu$ the dynamic viscosity), or $\frac{h_0^2}{12\mu}$ (with $h_0$ the gap height), respectively.
Substituting the continuity equation into the momentum equation yields
\begin{equation} \label{gov}
	\nabla\cdot\left(\alpha \nabla p\right)=0.
\end{equation}
For a homogeneous medium, Eq.~(\ref{gov}) reduces to the Laplace equation.

\begin{figure}[!ht]
	\centering
	\includegraphics[width=0.65\linewidth]{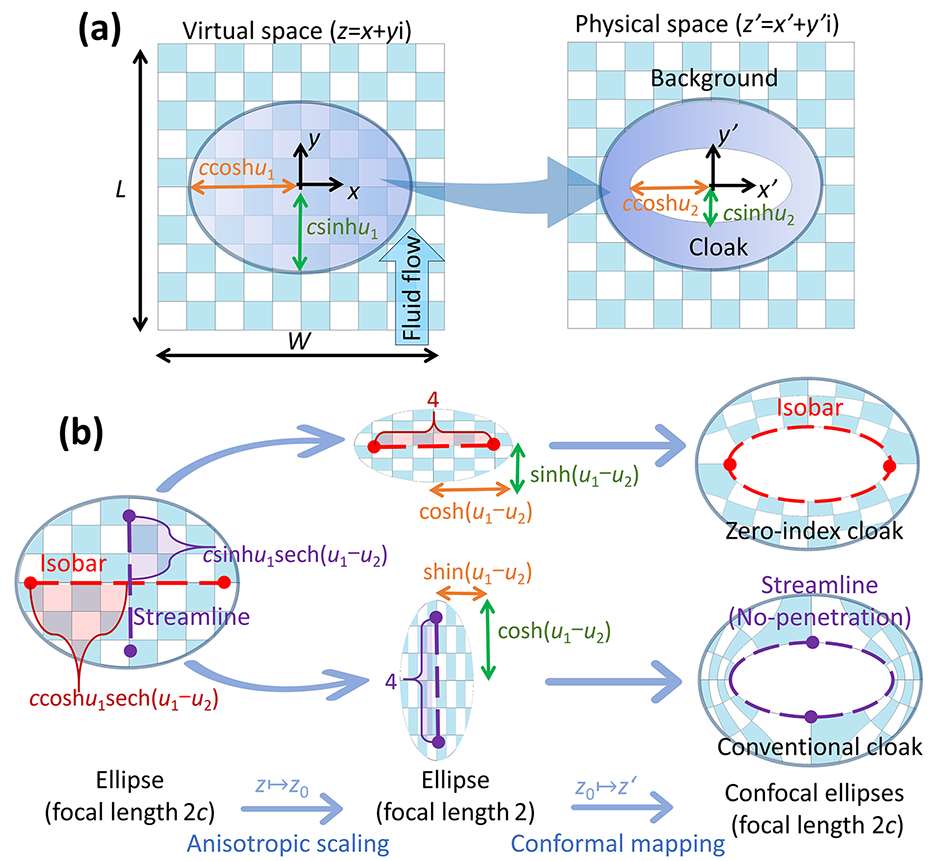}
	\caption{Pseudo-conformal mappings to design invisibility cloaks. (a) Transformation from the virtual space to the physical space. The coordinate origin is located at the center of the ellipse.
A small elliptical region inside a larger ellipse is removed by the transformation, forming a shell bounded by confocal ellipses. Their focal length is $2c$. (b) Two transformation paths differing in the major axis orientation of the intermediate ellipse. In both cases, the first step is an anisotropic scaling that produces an ellipse with focal length 2, followed by a conformal mapping.
In the upper path, a horizontal isobar in the virtual space becomes the inner boundary of the zero-index cloak. In the lower path, a vertical streamline becomes the inner boundary of the conventional cloak.}\label{1}
\end{figure}

\subsection{Pseudo-Conformal Mapping in hydrodynamics}

We design a cloak with confocal elliptical inner and outer boundaries using a two-step two-dimensional (2D) pseudo-conformal transformation~\cite{Dai2023b}: $z=x+y\text{i}\mapsto z_o=x_0+y_0\text{i} \mapsto z'=x'+y'\text{i}$. The Cartesian coordinates 
\((x, y)\), \((x', y')\), and \((x_0, y_0)\) refer to the virtual space, the physical space, and their intermediate, respectively.
 The major and minor axes of the outer ellipse are $c\cosh u_1$ and $c\sinh u_1$, respectively, while those of the inner ellipse are $c\cosh u_2$ and $c\sinh u_2$, i.e., their focal length is $2c$.
The preimage is an elliptical region with the same outer boundary. The first step is an anisotropic scaling along the $x$- and $y$-directions:
 \begin{equation}
	\left\{\begin{array}{l}
		x_0=x\dfrac{\cosh(u_1- u_2)}{\cosh u_1}, \\
		y_0=y\dfrac{\sinh(u_1- u_2)}{\sinh u_1},
	\end{array}\right.
\end{equation}
which transforms the elliptical region into another with focal length 2. This step is non-conformal. The second step is a conformal mapping: 
 \begin{equation}\label{iz}
	\left\{\begin{array}{l}z'=e^{ u_2}\frac{z_0-\sqrt{z_0^2-4}}{2}+\frac{1}{e^{ u_2}\frac{z_0-\sqrt{z_0^2-4}}{2}} \quad {\text{if }} x_0<0,\\ z'=e^{ u_2}\frac{z_0+\sqrt{z_0^2-4}}{2}+\frac{1}{e^{ u_2}\frac{z_0+\sqrt{z_0^2-4}}{2}}  \quad {\text{if }} x_0\geqslant 0,\end{array}\right.
\end{equation}
which tears the elliptical region along its focal axis and generates a shell enclosed by confocal ellipses. It can be verified that the combined mapping reduces to the identity transformation on the outer boundary of the cloak.

In conventional transformation optics, the parameter \(\alpha\) must be multiplied by the transformation factor \(\frac{\mathbf{J} \mathbf{J}^{\top}}{\det \mathbf{J}}\)~\cite{Pendry2006}. Here, $\mathbf{J}$ is the Jacobian of $z\mapsto z'$ and  $\det\mathbf{J}$ is its determinant.  This tensor factor introduces anisotropy in the physical space when the transformation is not conformal. 
However, when a pressure bias is applied along the \(y\)-direction in a rectangular virtual space of size \(L \times W\), the streamlines and isobars coincide with the \(y\)- and \(x\)-directions, respectively. The first-step transformation preserves the orthogonality of streamlines and isobars (i.e., being pseudo-conformal) and yields a transformation factor that is diagonal in this coordinate system:
$
\tanh u_1\coth(u_1 - u_2)\,\mathbf{e}_x\mathbf{e}_x + \coth u_1\tanh(u_1 - u_2)\,\mathbf{e}_y\mathbf{e}_y.
$
Here, \(\mathbf{e}_x\) and \(\mathbf{e}_y\) are the unit vectors in the \(x\)- and \(y\)-directions, respectively, respectively. They serve as the eigenvectors of the transformation tensor, and coincide with the directions of isobars and streamlines. Based on the conformality of the second step (Part A of Supplemental Material~\cite{SupMat}), in the physical space, streamlines and isobars remain the principal axes of \(\frac{\mathbf{J} \mathbf{J}^{\top}}{\det \mathbf{J}}\), and the corresponding eigenvalues remain unchanged at \(\coth u_1\tanh(u_1 - u_2)\) and \(\tanh u_1\coth(u_1 - u_2)\). 
Since only the component along the streamlines enters the governing equation, it is sufficient to retain that principal value alone, effectively eliminating anisotropy. Therefore, the  material parameter inside the cloak takes
\begin{equation}\label{tuo2}
	\alpha' = \alpha\,\coth u_1\tanh(u_1 - u_2) < \alpha.
\end{equation}
The construction above corresponds to the upper path in Fig.~~\ref{1}(b), where the major axes of all ellipses are aligned along the
$
x$-direction. In the virtual space, a selected isobar is mapped to the inner elliptical boundary of the cloak, which remains an isobar with the same pressure value in the physical space. Therefore, in addition to the material parameter inside the cloak, this design requires a prescribed boundary condition on the inner boundary.
Such a cloak is referred to as a zero-index cloak, whose inner boundary exhibits properties analogous to zero-index media in optics~\cite{Liberal2017,Kinsey2019} and thermotics~\cite{Li2019,Xu2020}: fluid can freely enter and exit, as if viscosity were absent. Meanwhile, the parameter 
$\alpha$ inside the cloak is smaller than that of the background, corresponding to either a lower permeability or higher viscosity, both impeding flow. These two effects balance each other to eliminate the scattering caused by both the cloaked region and the cloak itself.

Alternatively, flowing the lower path in Fig.~~\ref{1}(b), we modify the first step as
 \begin{equation}
	\left\{\begin{array}{l}
		x_0=x\dfrac{\sinh(u_1- u_2)}{\cosh u_1}, \\
		y_0=y\dfrac{\cosh(u_1- u_2)}{\sinh u_1},
	\end{array}\right.
\end{equation}
 such that the intermediate elliptical region has its major axis aligned along the 
$y
$-direction. The modified second step
\begin{equation}\label{iz2}
	\left\{\begin{array}{l}z'=e^{ u_2}\frac{z_0-\sqrt{z_0^2+4}}{2}+\frac{1}{e^{ u_2}\frac{z_0-\sqrt{z_0^2+4}}{2}} \quad {\text{if }} x_0<0,\\ z'=e^{ u_2}\frac{z_0+\sqrt{z_0^2+4}}{2}+\frac{1}{e^{ u_2}\frac{z_0+\sqrt{z_0^2+4}}{2}} \quad {\text{if }} x_0\geqslant 0\end{array}\right.
\end{equation}
 then produces the same cloak region. The material parameter $\alpha'$ inside the cloak is
 \begin{equation}\label{tuo1}
 	\alpha'=\alpha\coth u_1\coth( u_1- u_2)>\alpha,
 \end{equation}
 facilitating flow relative to the background. The inner elliptical boundary of the cloak now corresponds to a streamline. This implies a no-penetration boundary condition, where the velocity is strictly tangential. In earlier studies, this condition was realized by an inner layer with near-zero effective permeability, forming a conventional bilayer cloak~\cite{Tay2022,Chen2022b,Wang2021} when combined with the outer layer designed by Eq.~(\ref{tuo1}). However, in practice, flow around a solid obstacle is governed not by a no-penetration condition but by a no-slip condition, which prohibits even tangential velocity at the interface. This reveals an inherent inconsistency between the conventional cloak and viscous fluid flow. In contrast, the zero-index cloak avoids this issue. Of course, both cloaking strategies are still subject to viscous effects arising from no-slip boundaries at the sidewalls, which may cause deviations from the ideal potential flow model and should be carefully assessed.

\subsection{2D numerical verification}

 We perform 2D numerical  simulations via the finite-element software COMSOL Multiphysics. The entire  domain is a $L=120$~mm $\times$ $W=120$~mm square.
 The cloak parameters take $u_1=1,u_2=0.5$, and $c=30$~mm.
A unit pressure is applied along the $y$-direction. The background viscosity is $0.63$~Pa~s, comparable to that of glycerol.
 We conduct two sets of simulations, and each set compares the effects of the zero-index cloak and the conventional cloak. The first set employs the Darcy's law model consistent with the ideal potential flow model, while the second set utilizes a modified Hele-Shaw model described by
\begin{equation}\label{anna}
	\nabla p-\nabla \cdot\left(\mu \left(\nabla \mathbf{v} +(\nabla \mathbf{v}) ^{\top }\right)\right)+\frac{12\mu}{h_0^2}\mathbf v=0.
\end{equation}
This quasi-three-dimensional  model can be realized by activating the shallow channel approximation in the creeping flow module.
Unlike the ideal potential model, the modified model includes the viscous stress tensor $\mu \left(\nabla \mathbf{v} +(\nabla \mathbf{v})^{\top }\right)$ to reflect the finite width $W$ of the Hele-Shaw cell, with its effect becoming more prominent as $h_0$ increases.
For the Darcy's law, the permeability is set to $3\times 10^{-6}$~m$^{2}$.
For the modified Hele-Shaw flow, the gap height is $h_0=6$~mm, which results in the same effective permeability as Darcy's law. The boundary conditions for the two simulation sets are summarized in TABLE I. In both models, we modulate the viscosity to realize the cloak.

\begin{table}\label{t1}
	\caption{\label{tab:table1}Boundary conditions in 2D simulations [Fig.~\ref{2}].}
	\begin{ruledtabular}
		\begin{tabular}{ccccc}
			Model	&\multicolumn{2}{c}{Darcy's law}&\multicolumn{2}{c}{ Modified Hele-Shaw flow}\\
			Cloak type &Zero-index&Conventional&Zero-index
			&Conventional\\ \hline
			Inlet/Outlet&$y=\mp\frac{L}{2} $&$y=\mp\frac{L}{2}$ &$y=\mp\frac{L}{2}$&$y=\mp\frac{L}{2}$\\
			Sides ($x=\pm \frac{W}{2}$)&No-penetration&No-penetration
			&No-slip& No-slip\\	Inner boundary&Isobaric&No-penetration
			&Isobaric& No-slip\\
		\end{tabular}
	\end{ruledtabular}
\end{table}

\begin{figure}[!ht]
	\centering
	\includegraphics[width=0.9\linewidth]{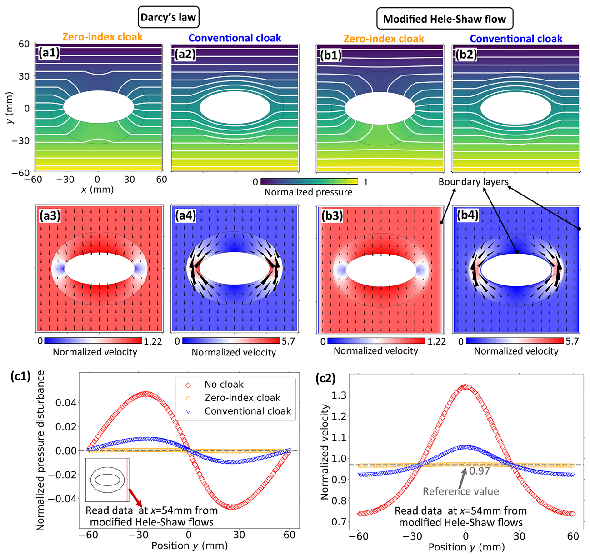}
	\caption{Simulation results of 2D cloaks when the pressure bias is applied along the $y$-direction. (a1), (a2)  Pressure distributions of (a1) the zero-index cloak and (a2) the conventional cloak, respectively, obeying Darcy's law. The white curves are isobars. (a3), (a4) The corresponding   velocity distributions under Darcy's law. The arrows indicate the velocity direction. (b1)–(b4) The same as (a1)–(a4), but for the modified Hele-Shaw model. (c1), (c2) The performance of cloaks under the modified Hele-Shaw model based on the (c1) pressure disturbance and (c2) velocity, respectively, on the line $x=54$~mm.}\label{2}
\end{figure}

Figure~\ref{2} presents the simulation results. Pressure value is normalized by the applied bias, and velocity by the uniform value in the virtual space under Darcy's law. For Darcy flow [Figs.~\ref{2}(a1)–\ref{2}(a4)], both cloaks achieve ideal cloaking performance. In contrast, under the modified Hele-Shaw model, boundary layers develop near the side walls [Figs.~\ref{2}(b3), \ref{2}(b4)], and an additional boundary layer near the inner boundary of the conventional cloak leads to noticeable bending of the isobars [Fig.~\ref{2}(b2)].
To quantify these differences, we examine the performance along the line \(x = 54\)~mm. In the virtual space, pressure varies linearly and velocity is uniform along this line~\cite{Note1}. Figs.~\ref{2}(c1) and \ref{2}(c2) show the pressure disturbance (i.e., deviation from the virtual-space value) and velocity profiles for both cloaks, as well as the case without a cloak. Although both cloaks significantly reduce the external-field distortion compared to the no-cloak case, the zero-index cloak exhibits closer agreement with the ideal solution because it is less affected by viscous forces.

In the above analysis, the pressure bias is applied along the \(y\)-direction. If the geometry is unchanged but the bias is applied along the \(x\)-direction, the parameter \(\alpha'\) for both cloaks must be modified, indicating that the cloaking effect is theoretically unidirectional. Part C of the Supplemental Material~\cite{SupMat} provides the parameter derivation and simulation results for this case, where the zero-index cloak still performs better than the conventional cloak.

\section{Three-dimensional omnidirectional  cloak via a hybrid strategy}

\subsection{Hybrid Cloaking Based on Conventional and Zero-Index Mechanisms}

\begin{figure}[!ht]
	\centering
	\includegraphics[width=0.77\linewidth]{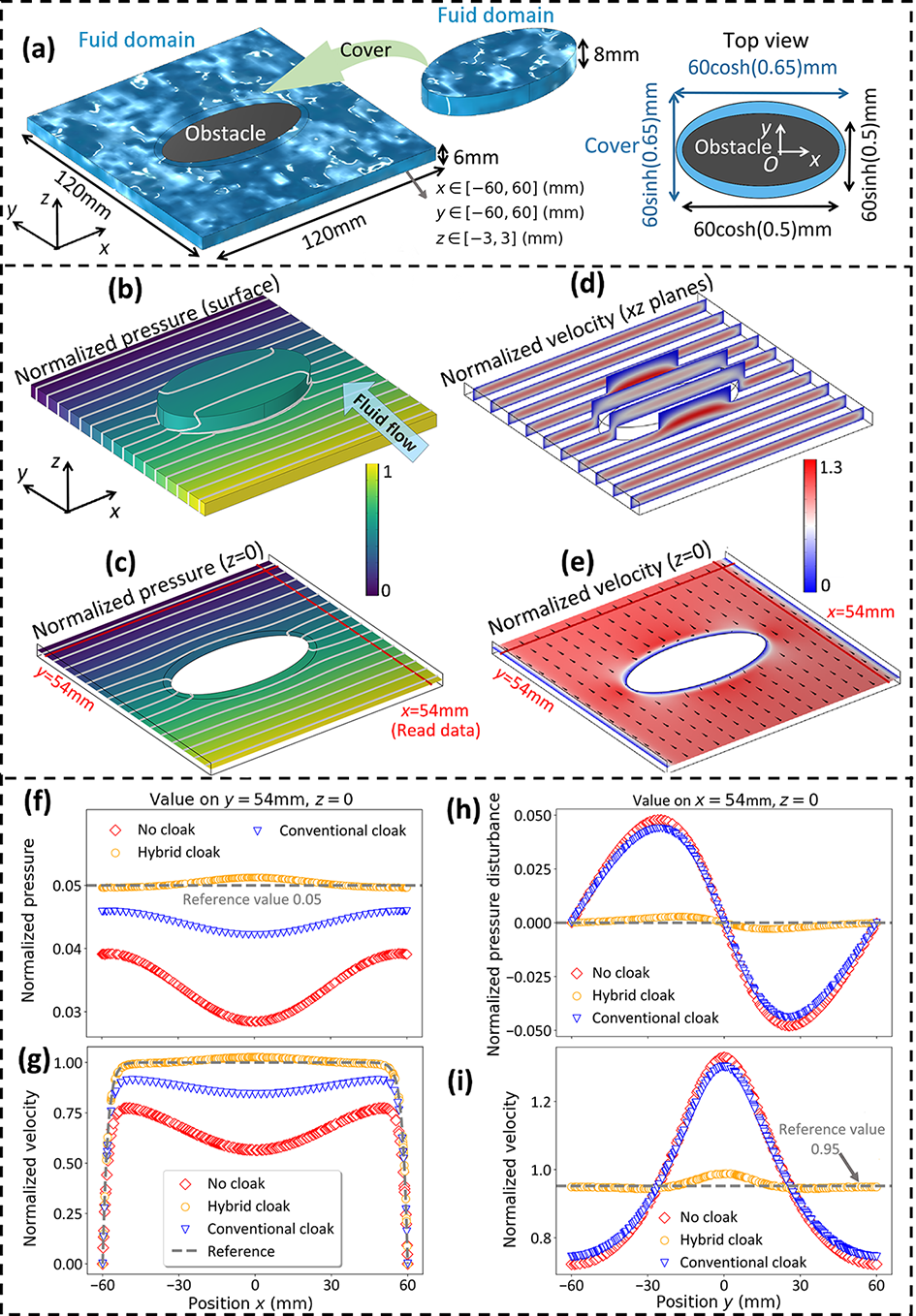}
	\caption{3D hybrid cloak. (a) Schematic of the hybrid cloak design. (b), (c) Simulated pressure distributions on (b) the top surface and (c) the central plane ($z = 0$). (d), (e) Simulated velocity distributions on (d) $xz$ planes and (e) the central plane ($z = 0$). (f)–(i) Comparison of cloaking performance. (f), (g) Pressure and velocity along $y = 54$~mm at $z = 0$. (h), (i) Pressure disturbance and velocity along $x = 54$~mm at $z = 0$.}
	\label{33}
\end{figure}

When considering three-dimensional (3D) fluid flow, the governing equation for Darcy’s law remains unchanged. In contrast, the Hele-Shaw model reverts to the full Stokes equation 
\begin{equation}\label{NSs}
	\nabla p-\nabla \cdot\left[\mu \left(\nabla \mathbf{v} +(\nabla \mathbf{v}) ^{\top }\right)\right]=0
\end{equation}
in a thin gap geometry. Part D of the Supplemental Material~\cite{SupMat} presents the design of both cloaks by adjusting the gap height inside the cloak. In the 3D case, the conventional cloak features a larger height than the background, while the zero-index cloak is shallower. Numerical results show that, similar to the 2D case, the zero-index cloak still outperforms the conventional cloak.
However, achieving the isobaric inner boundary condition remains a technical challenge. Streamlines indicate that one half of the inner boundary permits inflow, while the other allows outflow, forming an open boundary. One possible strategy is to actively regulate this flow using paired inlet and outlet pumps (e.g., forming dipoles~\cite{Jiang2025}), though such control would require external energy input and precise synchronization.

To avoid additional energy consumption, we connect the inflow and outflow through an upper channel that covers the top surface of the solid obstacle [Fig.~\ref{33}(a)]. This channel is shaped as an elliptical cylinder with a height denoted as the ``height of cover" \(h_c\). To form a continuous fluid domain, its cross-section must slightly exceed that of the obstacle. 
Intuitively, this protruding structure compensates for the space blocked by the obstacle, reducing the external-field distortion. As a result, it introduces two competing effects: the inner wall adds viscous resistance, while the elevated conduit promotes flow. Notably, the shallow layer originally used in the zero-index cloak becomes unnecessary.
We refer to this configuration as a ``hybrid cloak,'' as it integrates features from both conventional and zero-index designs. Compared to the conventional cloak, the hybrid cloak includes an additional fluid domain above the obstacle, while both share a larger gap height surrounding it.

Figure~\ref{33} shows the simulation results under a pressure bias $\Delta p$ along the $y$-direction. As before, pressure is normalized by the applied bias, and velocity by the uniform value in an ideal Hele-Shaw cell of infinite width, given by $\frac{h_0^2}{8\mu}\frac{\left| \Delta p \right|}{L}$.
The 3D simulations use the same viscosity ($0.63$~Pa·s), background gap height (6~mm), and planar geometry ($L=W=120$~mm, $u_2=0.5$, $c=30$~mm) as in the 2D case, but with a thinner cloak ($u_1=0.65$). The protrusion height is 8~mm; see Part~E of the Supplemental Material [Fig.~S5 and Table S3] for the parameter selection via seeking for the optimal height based on minimizing the external-filed distortion~\cite{SupMat}. As shown in Figs.~\ref{33}(b)–\ref{33}(e), the hybrid cloak maintains good performance in both pressure and velocity distributions.
To further evaluate the cloaking effect, we examine two cross-sectional lines at $z=0$: $y=54$~mm and $x=54$~mm [Figs.~\ref{33}(f)–\ref{33}(i)]. The former is theoretically an isobar (normalized to 0.05), showing boundary layers at both ends in velocity. The latter is a constant-speed line, where pressure should vary linearly.
Along both lines, the hybrid cloak closely matches the ideal reference, while the conventional cloak shows larger pressure deviations and velocity similar to the no-cloak case. The degraded performance of the conventional cloak arises from the thinner cloak requiring a greater cloak height, which increases the viscous effect. 
In Part~E of the Supplemental Material~\cite{SupMat}, we further demonstrate that an optimal height for the conventional cloak does not exist in most cases [Figs.~S6–S8]. We also examine the robustness of the chosen 8~mm protrusion height of the hybrid cloak by varying the cloak thickness and the background channel depth [Figs.~S7 and S8]. In addition, we discuss the selection of the hybrid cloak height under a pressure bias applied along the 
$x$-direction [Fig.~S9 and Table~S4].

\subsection{Evaluating the omnidirectional cloaking}

\begin{figure}[!ht]
	\centering
	\includegraphics[width=0.9\linewidth]{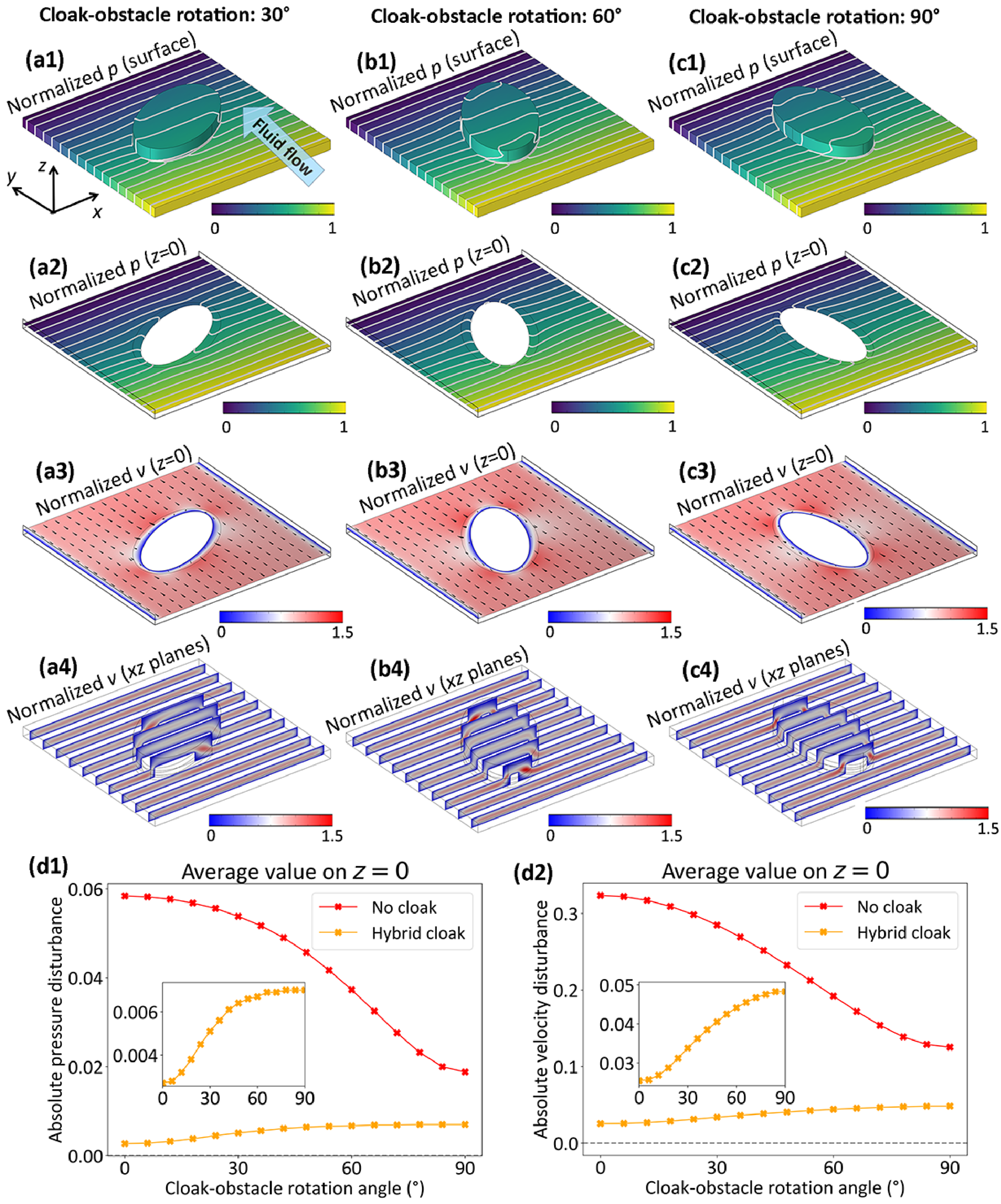}
	\caption{Simulation results of the hybrid cloak under different cloak–obstacle rotation angles.
		(a1)–(a4) The case for a rotation angle of 30°, including (a1) the pressure on the surface, (a2) the pressure on $z=0$, (a3) the velocity on $z=0$, and (a4) the velocity on $xz$ planes.
		(b1)–(c4) The same as (a1)–(a4), but for rotation angles of 60° and 90°, respectively.
		(d1), (d2) The average absolute (d1) pressure disturbance and (d2) velocity disturbance on $z=0$ as functions of the rotation angle. The subplots are magnified views of the data for the hybrid cloak.}\label{angle}
\end{figure}

The hybrid cloak differs from both the conventional and zero-index cloaks in that it is not constrained by the direction of the applied pressure bias. This allows us to evaluate its omnidirectional performance under an anisotropic geometry. Figure~\ref{angle} shows simulation results where the cloak and obstacle are rotated while the pressure bias remains fixed along the $y$-direction. Pressure and velocity fields are presented for rotation angles of 30$^\circ$ [Figs.~\ref{angle}(a1)–\ref{angle}(a4)], 60$^\circ$ [Figs.~\ref{angle}(b1)–\ref{angle}(b4)], and 90$^\circ$ [Figs.~\ref{angle}(c1)–\ref{angle}(c4)].
Figs.~\ref{angle}(d1) and \ref{angle}(d2) plot the deviations in pressure and velocity from perfect cloaking on the central plane, as a function of rotation angle (in 3$^\circ$ increments), averaged over the background region outside the cloak. Due to geometric symmetry, the range from 0$^\circ$ to 90$^\circ$ covers all possible orientations. When comparing velocity, we focus on the magnitude and ignore directional distortion. The results show that the hybrid cloak significantly suppresses background disturbance compared to the uncloaked case.
Further simulation details are provided in Part~F of the Supplemental Material~\cite{SupMat}, including cross-sectional data of Fig.~\ref{angle} [Fig.~S10] and quantitative evaluations of omnidirectional performance for various conventional and zero-index cloak configurations [Fig.~S11] . Overall, the hybrid cloak achieves the best omnidirectional performance.

\subsection{Experimental verification}

\begin{figure}[!ht]
	\centering
	\includegraphics[width=0.98\linewidth]{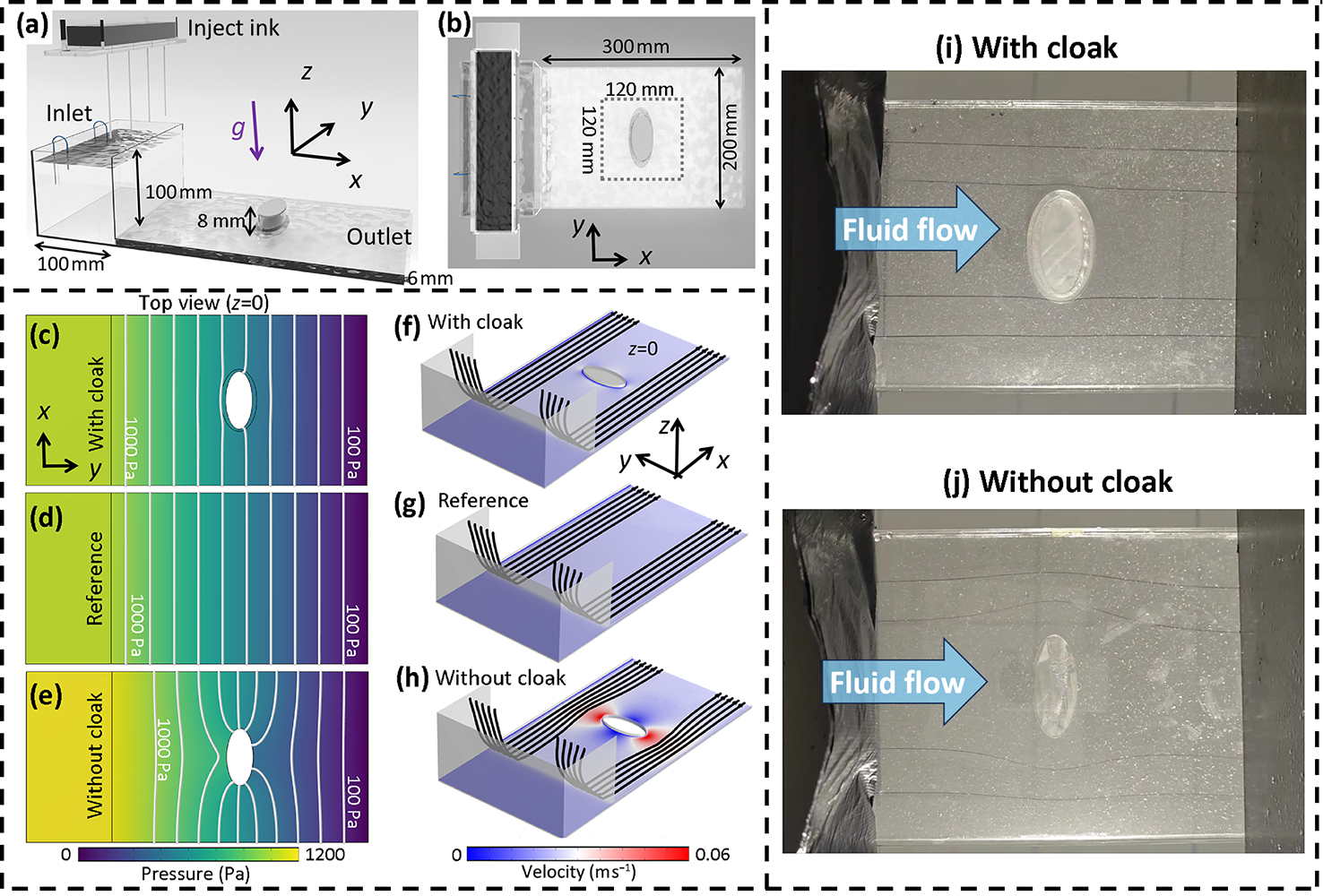}
\caption{Experimental demonstration. (a), (b) Schematics of the experimental setup: (a) side view and (b) top view. (c)–(e) Simulated pressure distributions at $z=0$ for (c) the hybrid cloak, (d) the reference case, and (e) the case without cloak. (f)–(h) Corresponding simulated velocity distributions at $z=0$ (projected onto the bottom surface) with simulated particle trajectories. (i), (j) Experimentally captured trajectories of ink particles: (i) with the hybrid cloak and (j) without cloak.} \label{55}
\end{figure}

The experimental setup is shown in Figs.~\ref{55}(a) and \ref{55}(b). The obstacle and hybrid cloak share the same structure as in Fig.~\ref{33}(a), and are placed at the center of a duct measuring $300~\mathrm{mm} \times 200~\mathrm{mm} \times 6~\mathrm{mm}$. An additional chamber ($100~\mathrm{mm} \times 200~\mathrm{mm} \times 100~\mathrm{mm}$) on the left is used to inject fluid (glycerol).
Due to gravitational acceleration $\mathbf{g}$, the realistic momentum equation is modified as
\begin{equation}\label{NS2}
	\nabla p - \nabla \cdot\left[\mu \left(\nabla \mathbf{v} + (\nabla \mathbf{v})^{\top} \right)\right] = \rho \mathbf{g},
\end{equation}
introducing a vertical pressure gradient. A fixed velocity is prescribed at the inlet, and the outlet is open to the atmosphere, where the static pressure is set to zero.
To visualize the streamlines, black ink particles are released from the inlet. Their trajectories are predicted using the Particle Tracing Module in COMSOL Multiphysics, with particle density and radius set to 1000~kg~m$^{-3}$ and $10^{-7}$~m, respectively. The simulation uses the “Newtonian, ignore inertial terms” option, along with Stokes drag, virtual mass, and pressure gradient forces. The injection velocity is 0.01~m~s$^{-1}$.

Figs.~\ref{55}(c)–\ref{55}(h) show the simulated pressure and velocity distributions on the $z=0$ plane. By comparing the white isobars, curvature of the black streamlines, and velocity magnitudes, the hybrid cloak still demonstrates effective cloaking even under gravity. The experimental particle trajectories—captured after reaching steady state—also align well with the simulations [Figs.~\ref{55}(i) and \ref{55}(j)].
Part~G of the Supplemental Material~\cite{SupMat} provides sample photographs and additional simulation details, including a tested case with a 90$^\circ$ rotation of the cloak-obstacle system. Since the disturbance reaches its maximum at 90$^\circ$ [Figs.~\ref{angle}(d1) and \ref{angle}(d2)], the structure demonstrates good omnidirectional cloaking performance.

\section{Discussion and Conclusion}
This study investigates hydrodynamic cloaking using transformation optics in combination with pseudo-conformal mapping. Hydrodynamic cloaking involves mapping a boundary-value problem in a fluid domain to that of another domain containing a hole. We emphasize that the boundary condition on the hole plays a critical role, although it has been underexplored in previous studies. The theoretical derivations are based on potential flow models, where transformation optics applies exactly. The cloak designs are further tested and optimized in Stokes flow between parallel plates, where the framework holds only approximately.
 We propose two distinct cloaking strategies via pseudo-conformal mappings, featuring contrasting boundary conditions on the inner surface, both realizable with isotropic and homogeneous media.
In real viscous flows, the no-slip condition on channel walls violates the irrotational assumption of potential flow. For conventional cloaks, imposing the no-slip condition on the inner boundary, deviating from the ideal no-penetration condition, also significantly degrades cloaking performance. The zero-index cloak leverages an isobaric condition at the inner boundary, offering a promising design strategy, albeit still idealized.
In both designs, the cloak parameters and the inner boundary conditions exert opposite effects on the flow field—one promoting and the other suppressing flow—to cancel disturbances in the background.
Motivated by this, we extend the flow-promoting isobaric condition into an additional flow passage above the obstacle. By integrating this with a raised fluid domain similar to the conventional cloak, we design a hybrid cloak with an anisotropic geometry.
In viscous flows, the hybrid cloak surpasses the conventional cloak in cloaking performance and shows better omnidirectionality than the zero-index cloak. Our conclusions are verified by numerical and experimental results.

The design strategy of the hybrid cloak offers two additional insights. First, by adjusting the vertical spacing between the upper and lower walls of the channel (i.e., introducing a grooved surface), we introduce a new degree of freedom for flow modulation, thereby avoiding the need for challenging viscosity engineering~\cite{Park2019,Sehgal2024}. This configuration remains effective even when fluids of different viscosities are introduced, without requiring structural changes to the cloak. Second, similar to elastodynamic cloaks~\cite{Yavari2019,Milton2006,Sozio2023}, hydrodynamic cloaks face the limitation that transformation optics does not strictly apply~\cite{Dai2023}. In our case, the height parameter of the hybrid cloak is obtained through numerical optimization. While analytical methods are valuable for guiding structural design, a paradigm shift toward more advanced optimization algorithms or machine learning techniques~\cite{Mirzakhanloo2020,Mirzakhanloo2020b,Ren2021,Wang2025b} is essential for future studies. Our approach and results can be applied in advanced microfluidic technologies, enabling functionalities such as particle manipulation~\cite{Khain2024,ZhouArxiv}, targeted drug delivery~\cite{Sanjay2018,Liu2022}, and selective cell sorting~\cite{Shields2015,Wang2023}. Furthermore, the pseudo-conformal mappings employed here can be directly extended to cloaking applications in other physical domains, including magnetostatics, electrostatics, and mass diffusion~\cite{Zhang2023,Yang2024}.

\section*{Acknowledgments}

We acknowledge the financial support from the National Natural Science Foundation of China under Grants No. 12305046, No. 12447112, and No. 12205101,  the Innovation Program of the Shanghai Municipal Education Commission under Grant No. 2023ZKZD06, the Postdoctoral Fellowship Program of CPSF
under Grant No. GZC20242261, and  the Shanghai Rising-Star Program under Grant No. 24QA2703100. Zhuo Li thanks the ``College Students' Innovation and Entrepreneurship Training Program" of Nantong University (No. 2024023).

\section*{Author contributions}
G.D. and Y.Z. contributed equally to this work. G. D., F.Y., and J.H. conceived the idea. G.D. Y.Z., and Z.L. conducted the theoretical calculations. G.D. and J.W. performed the numerical simulations. Y.Z., F.Y., and J. L. achieved the experimental observation and data analysis. All authors contributed to the discussion of the research, to the analysis of the results and to the writing of the manuscript.

\end{document}